\documentclass[%
 reprint,
 amsmath,amssymb, longbibliography,
 aps, 
 prd,
 twocolumn,
 showpacs,
 superscriptaddress
]{revtex4-2}

\usepackage{graphicx, placeins}
\usepackage{dcolumn}% Align table columns on decimal point
\usepackage{bm}% bold math
\usepackage[caption=false]{subfig}
\usepackage{float}
\usepackage[colorlinks=true, allcolors=blue]{hyperref}
\begin{document}

\preprint{APS/123-QED}

\title{Measurement of the differential cross section for neutral pion production in charged-current muon neutrino interactions on argon with the MicroBooNE detector}% Force line breaks with \\

% List of institutions in command form:
\newcommand{\ANL}{Argonne National Laboratory (ANL), Lemont, IL, 60439, USA}
\newcommand{\Bern}{Universit{\"a}t Bern, Bern CH-3012, Switzerland}
\newcommand{\BNL}{Brookhaven National Laboratory (BNL), Upton, NY, 11973, USA}
\newcommand{\UCSB}{University of California, Santa Barbara, CA, 93106, USA}
\newcommand{\Cambridge}{University of Cambridge, Cambridge CB3 0HE, United Kingdom}
\newcommand{\CIEMAT}{Centro de Investigaciones Energ\'{e}ticas, Medioambientales y Tecnol\'{o}gicas (CIEMAT), Madrid E-28040, Spain}
\newcommand{\Chicago}{University of Chicago, Chicago, IL, 60637, USA}
\newcommand{\Cincinnati}{University of Cincinnati, Cincinnati, OH, 45221, USA}
\newcommand{\CSU}{Colorado State University, Fort Collins, CO, 80523, USA}
\newcommand{\Columbia}{Columbia University, New York, NY, 10027, USA}
\newcommand{\Edinburgh}{University of Edinburgh, Edinburgh EH9 3FD, United Kingdom}
\newcommand{\FNAL}{Fermi National Accelerator Laboratory (FNAL), Batavia, IL 60510, USA}
\newcommand{\Granada}{Universidad de Granada, Granada E-18071, Spain}
\newcommand{\Harvard}{Harvard University, Cambridge, MA 02138, USA}
\newcommand{\IIT}{Illinois Institute of Technology (IIT), Chicago, IL 60616, USA}
\newcommand{\Indiana}{Indiana University, Bloomington, IN 47405, USA}
\newcommand{\KSU}{Kansas State University (KSU), Manhattan, KS, 66506, USA}
\newcommand{\Lancaster}{Lancaster University, Lancaster LA1 4YW, United Kingdom}
\newcommand{\LANL}{Los Alamos National Laboratory (LANL), Los Alamos, NM, 87545, USA}
\newcommand{\Louisiana}{Louisiana State University, Baton Rouge, LA, 70803, USA}
\newcommand{\Manchester}{The University of Manchester, Manchester M13 9PL, United Kingdom}
\newcommand{\MIT}{Massachusetts Institute of Technology (MIT), Cambridge, MA, 02139, USA}
\newcommand{\Michigan}{University of Michigan, Ann Arbor, MI, 48109, USA}
\newcommand{\MSU}{Michigan State University, East Lansing, MI 48824, USA}
\newcommand{\Minnesota}{University of Minnesota, Minneapolis, MN, 55455, USA}
\newcommand{\Nankai}{Nankai University, Nankai District, Tianjin 300071, China}
\newcommand{\NMSU}{New Mexico State University (NMSU), Las Cruces, NM, 88003, USA}
\newcommand{\Oxford}{University of Oxford, Oxford OX1 3RH, United Kingdom}
\newcommand{\Pitt}{University of Pittsburgh, Pittsburgh, PA, 15260, USA}
\newcommand{\Rutgers}{Rutgers University, Piscataway, NJ, 08854, USA}
\newcommand{\SLAC}{SLAC National Accelerator Laboratory, Menlo Park, CA, 94025, USA}
\newcommand{\SDSMT}{South Dakota School of Mines and Technology (SDSMT), Rapid City, SD, 57701, USA}
\newcommand{\Maine}{University of Southern Maine, Portland, ME, 04104, USA}
\newcommand{\Syracuse}{Syracuse University, Syracuse, NY, 13244, USA}
\newcommand{\TelAviv}{Tel Aviv University, Tel Aviv, Israel, 69978}
\newcommand{\Tennessee}{University of Tennessee, Knoxville, TN, 37996, USA}
\newcommand{\UTA}{University of Texas, Arlington, TX, 76019, USA}
\newcommand{\Tufts}{Tufts University, Medford, MA, 02155, USA}
\newcommand{\UCL}{University College London, London WC1E 6BT, United Kingdom}
\newcommand{\VTech}{Center for Neutrino Physics, Virginia Tech, Blacksburg, VA, 24061, USA}
\newcommand{\Warwick}{University of Warwick, Coventry CV4 7AL, United Kingdom}
\newcommand{\Yale}{Wright Laboratory, Department of Physics, Yale University, New Haven, CT, 06520, USA}
%%\newcommand{\listerThanks}{Now at University of Wisconsin, Madison}

% So that institutions appear in alphabetical order:
\affiliation{\ANL}
\affiliation{\Bern}
\affiliation{\BNL}
\affiliation{\UCSB}
\affiliation{\Cambridge}
\affiliation{\CIEMAT}
\affiliation{\Chicago}
\affiliation{\Cincinnati}
\affiliation{\CSU}
\affiliation{\Columbia}
\affiliation{\Edinburgh}
\affiliation{\FNAL}
\affiliation{\Granada}
\affiliation{\Harvard}
\affiliation{\IIT}
\affiliation{\Indiana}
\affiliation{\KSU}
\affiliation{\Lancaster}
\affiliation{\LANL}
\affiliation{\Louisiana}
\affiliation{\Manchester}
\affiliation{\MIT}
\affiliation{\Michigan}
\affiliation{\MSU}
\affiliation{\Minnesota}
\affiliation{\Nankai}
\affiliation{\NMSU}
\affiliation{\Oxford}
\affiliation{\Pitt}
\affiliation{\Rutgers}
\affiliation{\SLAC}
\affiliation{\SDSMT}
\affiliation{\Maine}
\affiliation{\Syracuse}
\affiliation{\TelAviv}
\affiliation{\Tennessee}
\affiliation{\UTA}
\affiliation{\Tufts}
\affiliation{\UCL}
\affiliation{\VTech}
\affiliation{\Warwick}
\affiliation{\Yale}

% Authors in alphabetical order
\author{P.~Abratenko} \affiliation{\Tufts}
\author{O.~Alterkait} \affiliation{\Tufts}
\author{D.~Andrade~Aldana} \affiliation{\IIT}
\author{L.~Arellano} \affiliation{\Manchester}
\author{J.~Asaadi} \affiliation{\UTA}
\author{A.~Ashkenazi}\affiliation{\TelAviv}
\author{S.~Balasubramanian}\affiliation{\FNAL}
\author{B.~Baller} \affiliation{\FNAL}
\author{G.~Barr} \affiliation{\Oxford}
\author{D.~Barrow} \affiliation{\Oxford}
% ...

\collaboration{The MicroBooNE Collaboration}
\thanks{microboone\_info@fnal.gov}\noaffiliation
%\email[]{microboone\_info@fnal.gov}

\date{\today}% It is always \today, today,

\begin{abstract}
We present a measurement of neutral pion production in charged-current interactions using data recorded with the MicroBooNE detector exposed to Fermilab’s booster neutrino beam. The signal comprises one muon, one neutral pion, any number of nucleons, and no charged pions. Studying neutral pion production in the MicroBooNE detector
provides an opportunity to better understand neutrino-argon interactions, and is crucial for future accelerator-based neutrino oscillation experiments. Using a dataset corresponding to $6.86 \times 10^{20}$ protons on target, we present single-differential cross sections in muon and neutral pion momenta, scattering angles with respect to the beam for the outgoing muon and neutral pion, as well as the opening angle between the muon and neutral pion. Data extracted cross sections are compared to generator predictions. We report good agreement  between the data and the models for scattering angles, except for an over-prediction by generators at muon forward angles. Similarly, the agreement between data and the models as a function of momentum is good, except for an underprediction by generators in the medium momentum ranges, $200-400$~MeV for muons and $100-200$~MeV for pions. %We report good agreement between the data and the models for scattering angles, whereas there are tensions between predicted and measured cross sections as a function of the momenta.
\end{abstract}

\maketitle
\raggedbottom
%\tableofcontents
\setlength{\parskip}{0pt}
\section{Introduction}
Studying neutrino-nucleus interactions is crucial in addressing longstanding fundamental questions in neutrino physics~\cite{PhysRevD.98.030001,PhysRevLett.123.151803,Nature}. Multiple current and future accelerator-based neutrino experiments use liquid argon time projection chambers (LArTPCs) as the detection technology~\cite{doi:10.1146/annurev-nucl-101917-020949,DUNE:2020jqi}. Therefore, a precise understanding of neutrino-nucleus interactions in argon is critical to optimizing the physics program of future experiments, such as the Deep Underground Neutrino Experiment (DUNE)~\cite{DUNE:2020jqi}. In neutrino oscillation experiments that measure electron-neutrino appearance rates in a muon-neutrino beam, a dominant background arises from neutral pions ($\pi^0$) decaying to two photons with only one photon being reconstructed successfully. Similarly, $\pi^0$ photons are often a dominant background in Beyond the Standard Model searches which target photon or electron-positron final-states~\cite{PhysRevLett.128.111801, PhysRevD.106.092006}. Hence, accurate modeling of $\pi^0$ production in charged-current~(CC) and neutral-current (NC) interactions enables precise predictions of background rates.

In this work, we present the first differential cross section measurement of $\nu_{\mu}$ CC interactions on argon with neutral pions in the final state. The event topology contains a muon, a single $\pi^0$ meson, any number of nucleons, and no charged pions in the final state,%The interaction final states in this measurement are defined as
\begin{equation}
    \nu_{\mu} + \text{Ar} \rightarrow \mu^- + \pi^0 + 0 \pi^{\pm} + \text{X},
\end{equation}
where $\text{Ar}$ represents the struck argon nucleus, $\text{X}$ represents the residual nucleus and any number of ejected protons or neutrons, but no other hadrons or leptons. We refer to these events as $\nu_{\mu}$ CC1$\pi^0$. 
This interaction commonly occurs through the $\Delta(1232)$ resonance for neutrinos with energy below $2$ GeV. There is no coherent contribution to this process since the final state for CC coherent pion production includes a charged pion. This interaction is therefore an ideal probe of incoherent processes that can constrain models of neutral pion production. 

We report the single differential cross section in muon and $\pi^0$ momenta, scattering angles with respect to the beam for the outgoing muon and $\pi^0$, as well as the opening angle between the muon and $\pi^0$. In order to improve the efficiency and probe a larger phase space of kinematic variables, the event selection accepts muon candidates that are either contained or exiting the detector volume. This measurement is not only relevant to MicroBooNE, but also to future LArTPC neutrino experiments such as the short baseline neutrino (SBN) program~\cite{doi:10.1146/annurev-nucl-101917-020949} and DUNE. 

The total flux-integrated cross section of $\nu_{\mu}$ CC single $\pi^0$ production on argon has been reported by the MicroBooNE collaboration~\cite{PhysRevD.99.091102}. Previous single-differential measurements of $\pi^0$ production in CC neutrino interactions were performed on nuclei lighter than argon. The $\nu_{\mu}$ $\textrm{CC}1\pi^0$ cross section was measured on carbon in the MiniBooNE experiment in $2011$~\cite{PhysRevD.83.052009}, and in the MINERvA experiment in $2015$~\cite{LE2015130, PhysRevD.96.072003}. A measurement of the $\nu_{\mu}$ $\textrm{CC}1\pi^0$ cross section on water has been reported by the K2K collaboration, presented as a ratio to the (CCQE) cross section~\cite{PhysRevD.83.054023}. The latest results by the NOvA collaboration use a more inclusive signal definition~\cite{NOVA_LATEST_PRD}. 
Measuring the $\pi^0$ kinematic distributions in argon can be used to benchmark the final state interaction modeling (which increase with mass number) used in event generator simulations as well as for testing the validity of resonance models.
\section{MicrobooNE Experiment}
MicroBooNE is an $85$ metric ton LArTPC on the booster neutrino beam (BNB) at Fermilab~\cite{Acciarri_2017}.  The dimensions of the MicroBooNE detector are $2.56 \times 2.32 \times 10.35$ m$^3$. The beam reaching the detector has a mean energy of $0.8$ GeV and is predicted to contain $93.6$\% muon neutrinos~\cite{osti_1212167}. The pion production measurement employs the muon neutrino component of the beam using data collected from $2016-2018$, which corresponds to $6.86\times10^{20}$ protons on target (POT). Charged particles traversing the liquid argon volume of the detector produce ionization and create prompt ultraviolet scintillation light. An electric field of $273$ V/cm is applied between the cathode and anode planes. With the applied electric field of the TPC, the ionization electrons drift horizontally towards the anode planes and are detected by wires in two induction planes. The charge is then deposited on the collection plane wires. The collection plane wires are oriented vertically, and the induction plane wires are oriented at angles $\pm 60^{\circ}$ with respect to the vertical direction. A light detection system with $32$ photomultiplier tubes detects the scintillation photons. Information from the three wire planes and the light detector can be combined to derive 3D images of the path of charged particles in the TPC. 
\section{Simulation Synopsis}
The framework developed by the MiniBooNE collaboration is leveraged to simulate the neutrino flux at the MicroBooNE detector~\cite{PhysRevD.79.072002}. Neutrino-nucleus interactions are simulated using the GENIE $\text{v3.0.6 G18\_10a\_02\_11a}$~\cite{GENIEV3} event generator prediction with a custom tune developed using T2K data~\cite{PhysRevD.105.072001} for the MicroBooNE analyses~\cite{PhysRevLett.128.151801}. This includes generating the primary neutrino interaction within the nucleus, producing all final state particles, and the interactions of the final state particles through the nucleus. The tune modifies the default GENIE $\text{v3.0.6 G18\_10a\_02\_11a}$ prediction for quasi-elastic (QE) and meson exchange current (MEC) models, but has no effect on resonant (RES) interactions. The Berger-Sehgal model is used for resonant pion production in the above mentioned version of GENIE~\cite{10.1063/1.3274164,PhysRevD.76.113004,PhysRevD.79.079903}. 

The simulation of the detector response starts with Geant4~\cite{AGOSTINELLI2003250} for particle propagation, and continues with LArSoft~\cite{Snider_2017} for simulating the anode wire signals and the scintillation light in the PMTs. A lookup table from a Geant4 photon propagation simulation is used to model the scintillation light response. The electric field distortions resulting from space charge effects are incorporated using data-driven electric field maps~\cite{Adams_2020,Abratenko_2020}. A modified box model is employed to simulate the ion recombination~\cite{R_Acciarri_2013}. The drift electron lifetime and the wire response is modeled with a time dependent simulation~\cite{Adams_2018_1, Adams_2018_2,R_Acciarri_2013}. Because of its near-surface location, the MicroBooNE detector is exposed to a significant amount of cosmic rays resulting as backgrounds. Cosmic ray events recorded during off-beam data taking are used to estimate the background arising from such events. The event selection requirement (discussed in~\ref{evt_sel}) is applied to this off-beam data to estimate the background. To model the background from cosmic rays in neutrino-induced triggers, we overlay unbiased data collected in a beam-off environment onto a simulated neutrino interaction. With this approach, the detector noise is also incorporated in a data-driven manner. %With this approach, the detector noise is also incorporated in a data-driven manner. With this approach, the detector noise is also incorporated in a data-driven manner. 
\section{Analysis Overview}
The reconstruction chain starts with noise removal~\cite{Noise_removal} and signal processing~\cite{Adams_2018_2}. The Pandora pattern recognition toolkit is used to reconstruct neutrino candidate events~\cite{R_Acciarri_2018_Pandora}. A neutrino event is selected by rejecting cosmic rays crossing the detector and leveraging optical information coincident with the beam window. Particles such as electrons and photons leave a shower-like signature, while protons, muons, and charged pions leave track-like signatures within the detector. MicroBooNE’s log-likelihood ratio particle identification tool is used to achieve better particle identification performance within track-like objects to classify muons and protons~\cite{JHEP}. For tracks contained within the detector volume, energies are estimated using the track range~\cite{148751}, while multiple Coulomb scattering is used for exiting tracks~\cite{Abratenko_2017, LYNCH19916, HIGHLAND1975497}. For electromagnetic showers, calorimetric energy reconstruction is performed by summing the total clustered energy deposits within the shower. A correction factor of $1.2$ is applied to account for the inefficiencies attributed to clustering of charge~\cite{PhysRevD.105.112004}.

We define the signal in terms of the observable final state particles. Hence, all truth-level event definitions mentioned in this analysis only consider particles produced after the final state interactions. A neutrino scattering event is chosen as part of the signal if it contains a muon with kinetic energy greater than $20$ MeV, exactly one $\pi^0$ meson, no charged pions with kinetic energy greater than $40$ MeV, and any number of nucleons. The kinetic energy thresholds are driven by reconstruction efficiencies~\cite{supp_mat}. 
We describe the major event selection strategies in the following subsection. 
\subsection{Event Selection \label{evt_sel}}
The signal consists of muon-neutrino induced $\pi^0$ production in a CC interaction within the fiducial volume, where the $\pi^0$ meson decays to photons. Hence, the event selection focuses on identifying $\nu_{\mu}$ CC events associated with two photon showers from the $\pi^0$ candidate, a contained or exiting muon track, any number of nucleons, and no charged pions. We require the neutrino interaction vertex to be within the fiducial volume (FV). For this measurement the FV is chosen to be $10$ cm from the sides of the TPC along the drift direction, $15$ cm from the sides of the TPC in the vertical direction, $10$ cm from the upstream face, and $50$ cm from the downstream face of the TPC with respect to the beam direction. The selection criteria accept signal events with muon candidate tracks either contained or exiting the FV. All muon candidate tracks (contained and exiting) are required to have a Pandora track score greater than $0.5$. The track score assigned by Pandora classifies reconstructed particles as track-like (scores closer to 1) or shower-like (scores closer to 0). For the contained muon candidate tracks, we require that the track's start and end coordinates be within the FV, the track length be $> 10$ cm, and the candidate track start within $3$ cm of the reconstructed neutrino interaction vertex. The log-likelihood ratio particle identification (LLR PID) score~\cite{JHEP} is required to be $> 0.2$. The LLR PID discriminates particle tracks by assigning scores ranging from $-1$ to $1$, with $-1$ being most proton-like and $1$ being most muon-like. For the exiting muon candidate tracks, the selection criteria differ in the containment and track length requirements. Here, we require the start of the track to be within the FV, along with a selection criteria that the track length be $> 30$ cm. Additionally, the selection criteria require that the number of charged pions be zero. The track signature of charged pions within the LArTPC is similar to that of muons. The charged pion veto is implemented by constraining the number of such tracks to exactly one. This selects one muon candidate track. 
After selecting events consistent with the signal topology, a set of selection criteria is applied to reject backgrounds. These requirements include events that have the most energetic shower above a $40$ MeV energy threshold with the cosine of the radial angle greater than $0.9$. Additionally, the conversion distance of the most energetic shower is required to be less than $80$~cm and events with smaller conversion distance ($< 2$~cm) are selected only if the energy deposition per unit length ($dE/dX$) is $>2.5$~MeV/cm.  The conversion distance refers to the distance between the reconstructed interaction vertex and the shower start point, and the radial angle is the angle between the shower direction vector and the vector connecting the neutrino interaction vertex and the shower start point. The requirement on the small conversion distance in combination with $dE/dX$ reduces contribution from tracks that are mis-reconstructed as showers near the interaction vertex, whereas rejecting events with higher conversion distance ($> 80$~cm) removes random coincidence with cosmic events. 
\begin{figure*}
    \subfloat[\label{sfig:pi0mom}]{
  \includegraphics[width=0.47\textwidth]{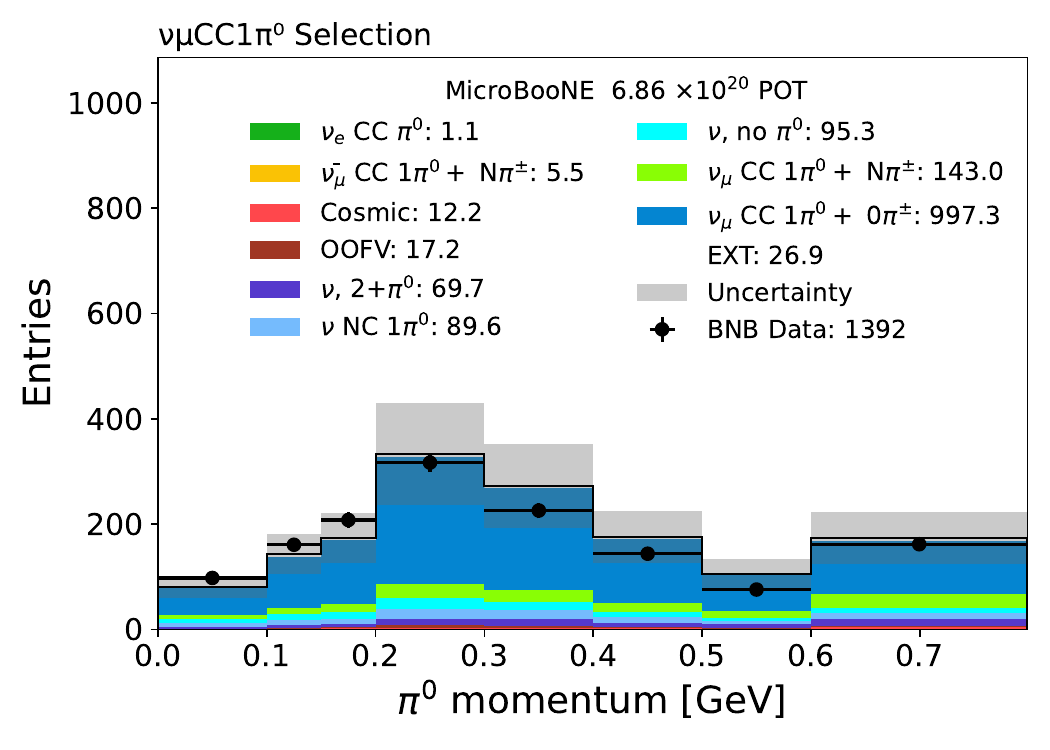}}\hfill
  \subfloat[\label{sfig:pi0angle}]{
  \includegraphics[width=0.47\textwidth]{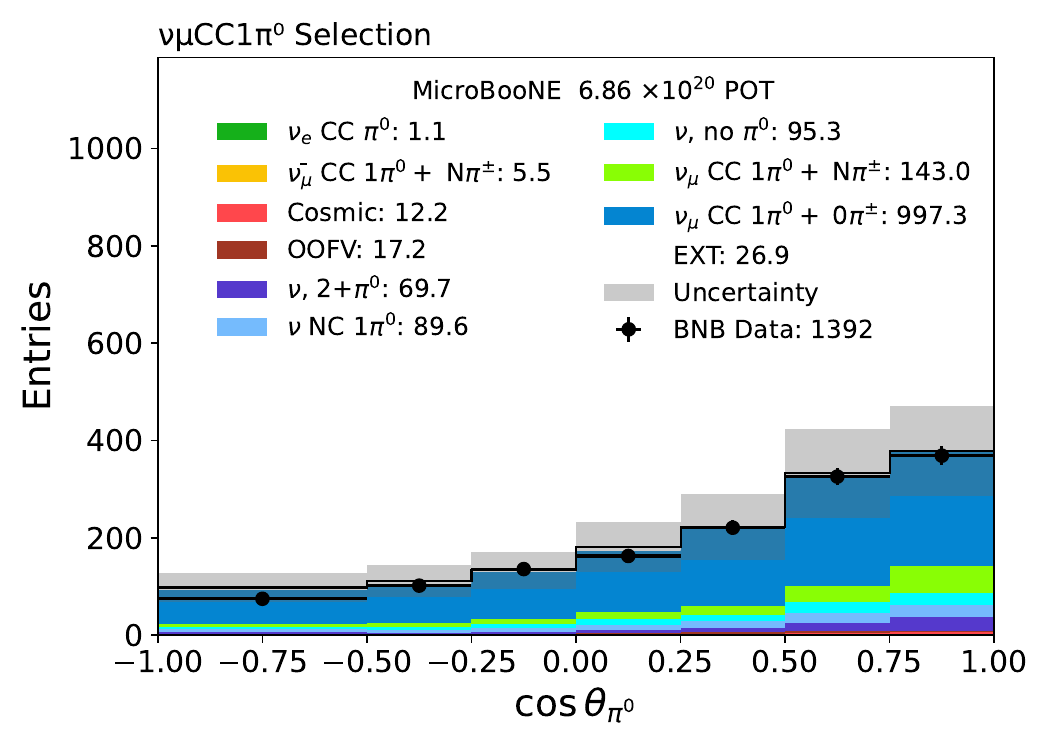}} \\ \vspace{1.5mm}
  \subfloat[\label{sfig:mumom}]{
  \includegraphics[width=0.47\textwidth]{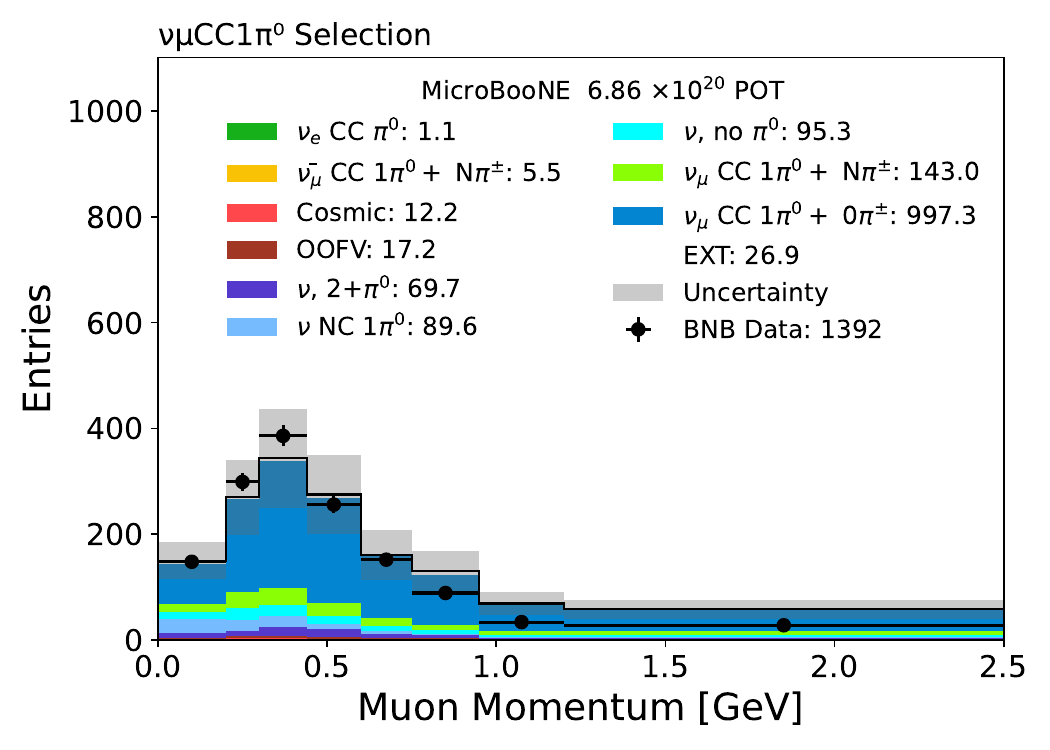}}\hfill
  \subfloat[\label{sfig:muangle}]{
  \includegraphics[width=0.47\textwidth]{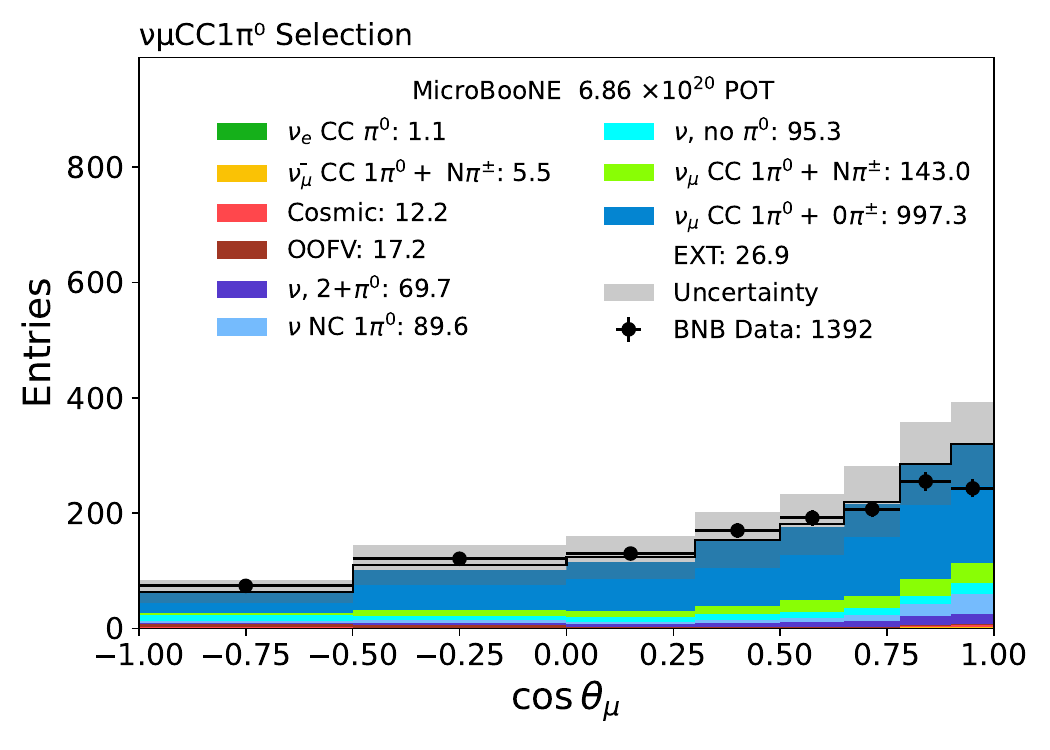}} \\ \vspace{1.5mm}
  \subfloat[\label{sfig:pimuangle}]{
  \includegraphics[width=0.457\textwidth]{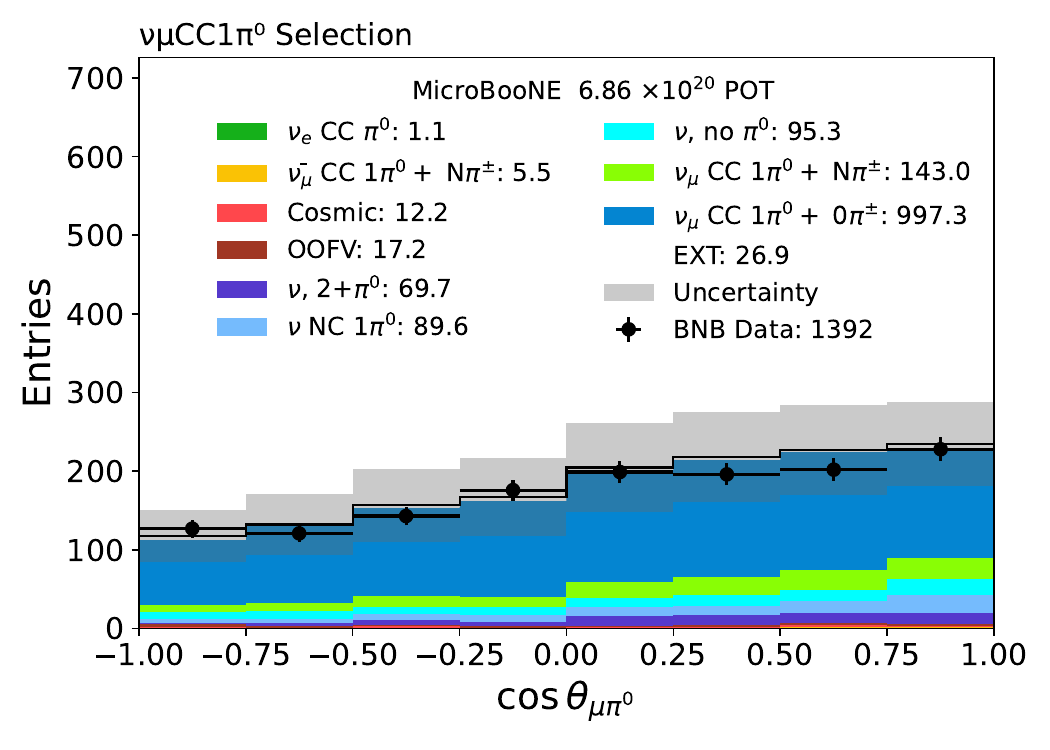}} 
\caption{ 
    \label{sfig:dist} Comparison of the distribution of the number of events observed in data and the prediction of the MicroBooNE tuned version of GENIE v3 satisfying the $\nu_{\mu}$CC$1\pi^0$ selection requirements for  (a) the $\pi^0$ momentum, (b) the scattering angle between the neutrino beam and outgoing $\pi^0$ direction, (c) the muon momentum, (d) the scattering angle between the neutrino beam and outgoing muon direction, and (e) the opening angle between the muon and pion are shown. The last bin in the momentum distributions acts as an overflow bin. The shaded gray band indicates the total (includes both statistical and systematic) uncertainty on the simulation prediction. The event categories are described in the legend. The OOFV category includes background events in which the true neutrino vertex is located outside the fiducial volume.}
\end{figure*}
The selection criteria for sub-leading showers starts with an energy threshold of $10$~MeV. Additionally, the sub-leading shower conversion distance is required to be $> 1$~cm with a similar approach on selecting events with smaller values ($< 1$~cm) using $dE/dX$ ($> 2.5$~MeV/cm) information as implemented for the leading showers. The final requirement is on the reconstructed $\pi^0$ invariant mass to be within $50 - 180$~MeV. The total signal selection efficiency is $8.5$\% and the purity is $69$\%. 

The predicted contribution of background events in the selection is about $31$\%. A large fraction of the backgrounds that pass the event selection criteria arises from CC $\pi^0$ events containing $\pi^{\pm}$ or NC events containing $\pi^0$ mesons in the final state. The second largest background contribution comes from neutrino interactions with no final state $\pi^0$ meson. These events can result from scenarios where a final state charged pion produces a $\pi^0$ meson as a result of re-interaction with another nucleus. The distributions from which the differential cross sections will be unfolded are shown in Fig.~\ref{sfig:dist}. The predicted event categories from the MicroBooNE tuned version of GENIE $\text{v3.0.6}$ $\text{G18\_10a\_02\_11a}$ satisfying the $\nu_{\mu}$ CC$1\pi^0$ selection criteria are shown as a stacked histogram and are compared with the distribution of the number of events in data. In total, $1392$ events from the data sample satisfy the $\nu_{\mu}$ CC$1\pi^0$ selection criteria. The predicted number of events from simulation agrees well with the data for outgoing muons and pions except for muons with high momenta. The outgoing muon kinematics is subject to nuclear structure. Comparing the cross section as a function of kinematic variables extracted from data with those predicted by the nuclear structure models that enter the event simulations can provide insight into the validity of these models. The outgoing $\pi^0$ kinematics are sensitive to final state interactions. This is apparent in the MINERvA collaboration’s latest CC$\pi^0$ cross-section measurement on carbon~\cite{MINERvA:2020anu}. About $83$\% of the predicted events come from RES interactions, followed by a $14.3$\% contribution from deep inelastic scattering (DIS) interactions~\cite{supp_mat}. Several sources of uncertainties are taken into account while reporting the total number of predicted events and are described next. %A detailed description of the systematic uncertainty evaluation is provided in the next section.
\subsection{Systematic Uncertainties\label{sys_unc}}
The systematic uncertainties for this measurement come from the BNB neutrino-flux prediction, variations in modeling of neutrino interactions and final state re-interactions, and variation of detector simulation parameters. 

The neutrino flux uncertainties arise from hadron production uncertainties after the initial proton hits the target, the beamline uncertainties related to horn current, and position of the target~\cite{PhysRevD.79.072002}. These are taken into account by applying a reweighting scheme to the simulation and are approximately at the $7-10$\% level. 
The systematic uncertainties for generating neutrino interaction events within GENIE are taken into account. The interaction models are computed by reweighting the MicroBooNE tune GENIE model parameters~\cite{PhysRevD.105.072001}. Cross section model uncertainties on signal events are incorporated by evaluating the effect of model variations on the smearing and efficiency of the predicted event rate, and on the rate of predicted neutrino backgrounds. The outgoing particles from the primary neutrino interaction may interact with other nuclei inside the detector. This effect is modeled by a similar reweighting approach~\cite{Calcutt_2021}. The reinteraction uncertainties are approximately at the $2$\% level and have small impact on the results reported. 

The detector systematic uncertainties are estimated using dedicated samples that are generated by changing model parameters in the nominal simulation~\cite{det_unc}. Several effects of the detector response model have been considered such as wire response, space charge, electron-ion recombination, and light yield. The uncertainties arising from detector systematic effects for the $\nu_{\mu}$ CC1$\pi^0$ measurement are at the $5-15$\% level across all five kinematic variables. Additional normalization uncertainties at the $2$\% and $1$\% level come from the POT counting and the estimated number of argon nuclei in the detector respectively. The flux and detector effects are the dominant sources of uncertainties in the $\nu_{\mu}$ CC1$\pi^0$ measurement.  

As the differential cross section is measured across multiple correlated bins, we utilize a covariance matrix to incorporate the uncertainties in the final calculation of cross sections. The covariance matrix for each source of systematic uncertainty is determined by considering multiple systematic variations and is given by
\begin{equation}
    V_{ij} = \frac{1}{N}\sum_{k=1}^N (n_i^{\textrm{CV}} - n_i^k)(n_j^{\textrm{CV}} - n_j^k),
\end{equation}
where the elements of the covariance matrix $V_{ij}$ are expressed in terms of a sum over all of the systematic variations. Here, $N$ represents the total number of systematic variations, $n_i^{\textrm{CV}}$ the number of reconstructed events in bin $i$ for the central-value (CV) simulation, and $n_i^k$ is a prediction of the same quantity when a systematic variation is applied to the CV sample.  
The total covariance matrix 
\begin{equation}
V_{\textrm{tot}} = V_{\textrm{sys}} + V_{\textrm{stat}}
\end{equation}
encodes the systematic uncertainties and data statistical uncertainties, where $V_{\textrm{sys}}$ is the sum of the covariance matrices from all of the systematic contributions discussed above, and $V_{\textrm{stat}}$ consists of an uncorrelated diagonal statistical covariance matrix.
\begin{figure*}
    \subfloat[\label{sfig:pi0momxsec}]{
  \includegraphics[width=0.45\textwidth]{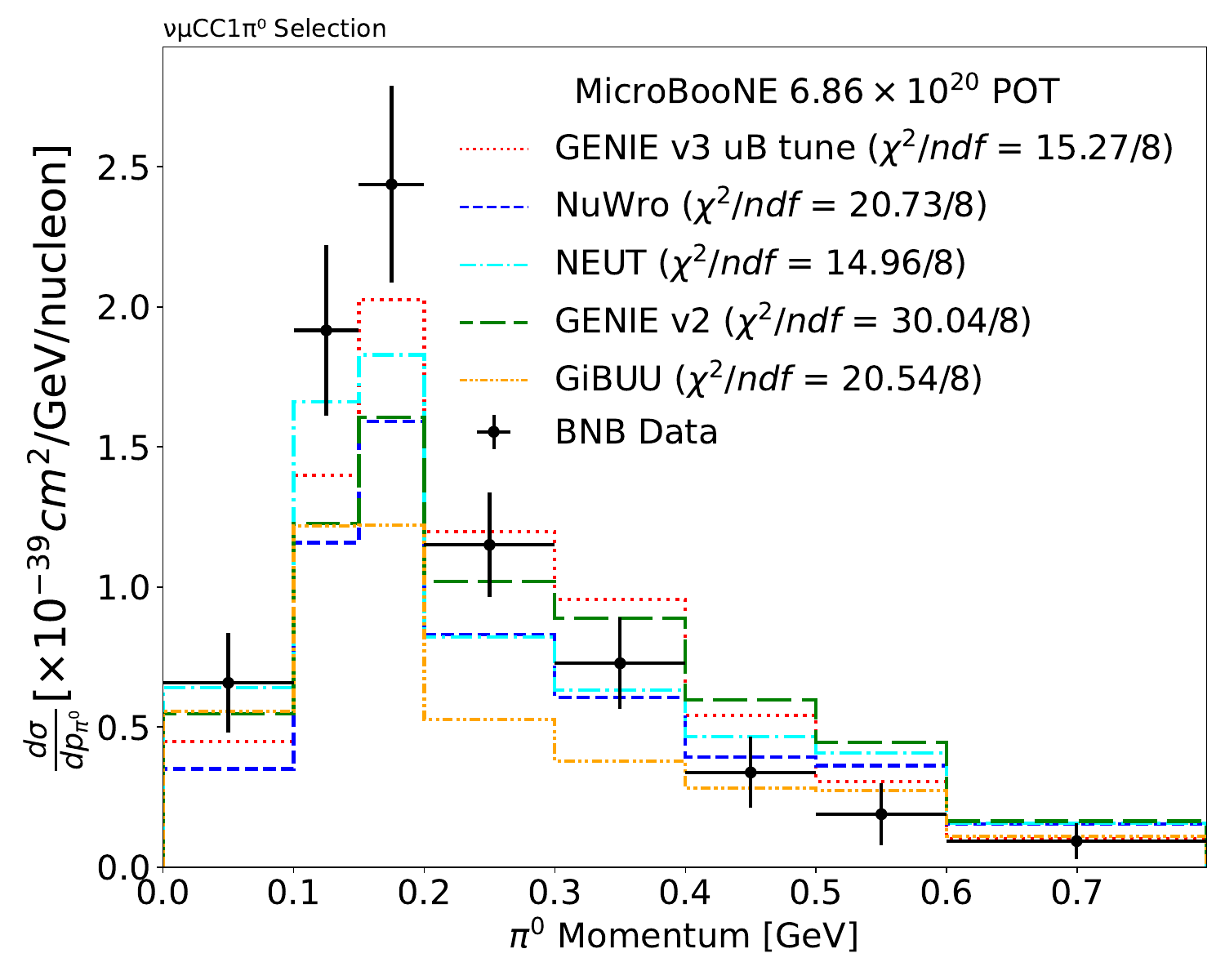 }}\hfill
    \subfloat[\label{sfig:pi0anglexsec}]{
  \includegraphics[width=0.45\textwidth]{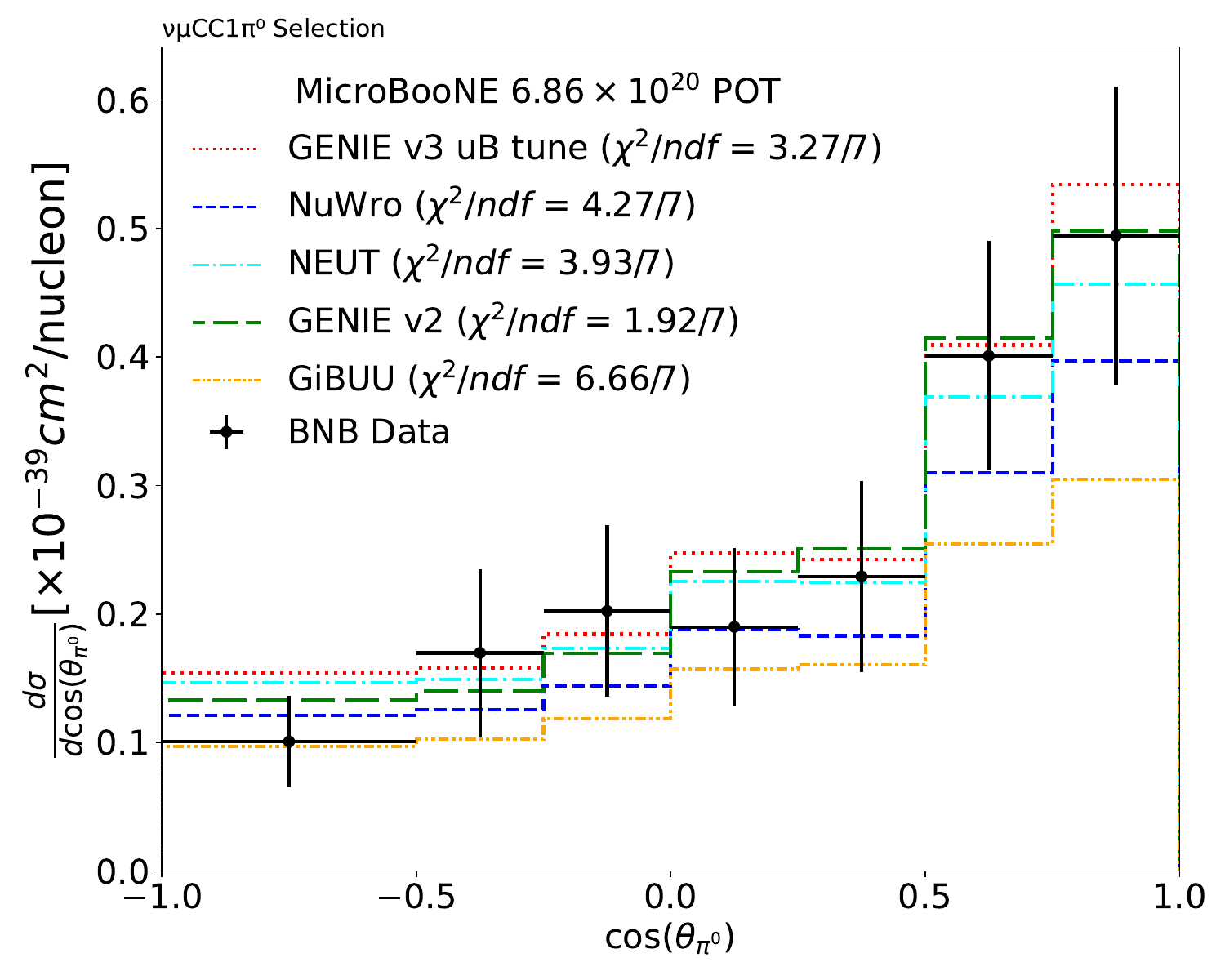}}\hfill
    \subfloat[\label{sfig:mumomxsec}]{
  \includegraphics[width=0.45\textwidth]{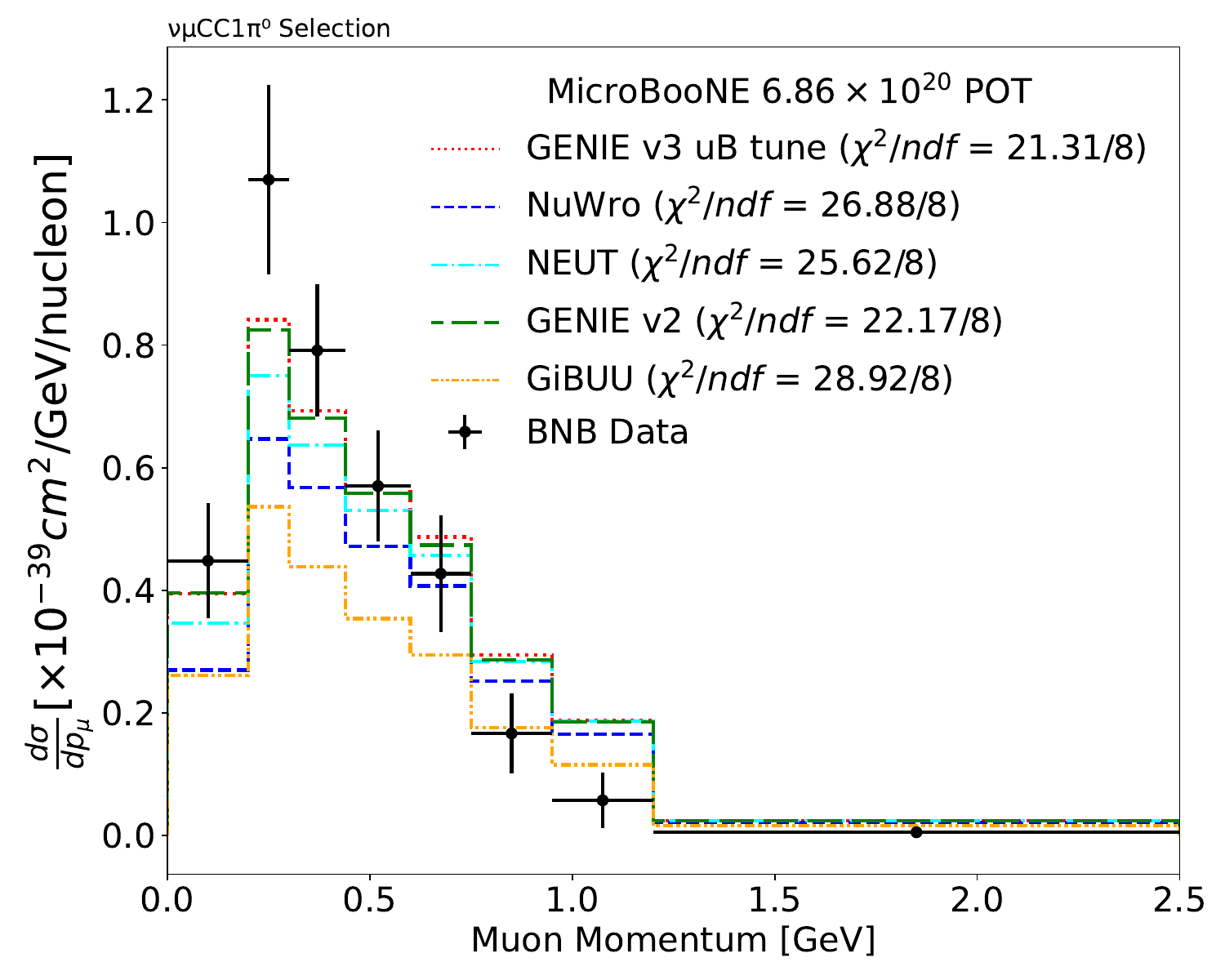}}\hfill
  \subfloat[\label{sfig:muanglexsec}]{
  \includegraphics[width=0.45\textwidth]{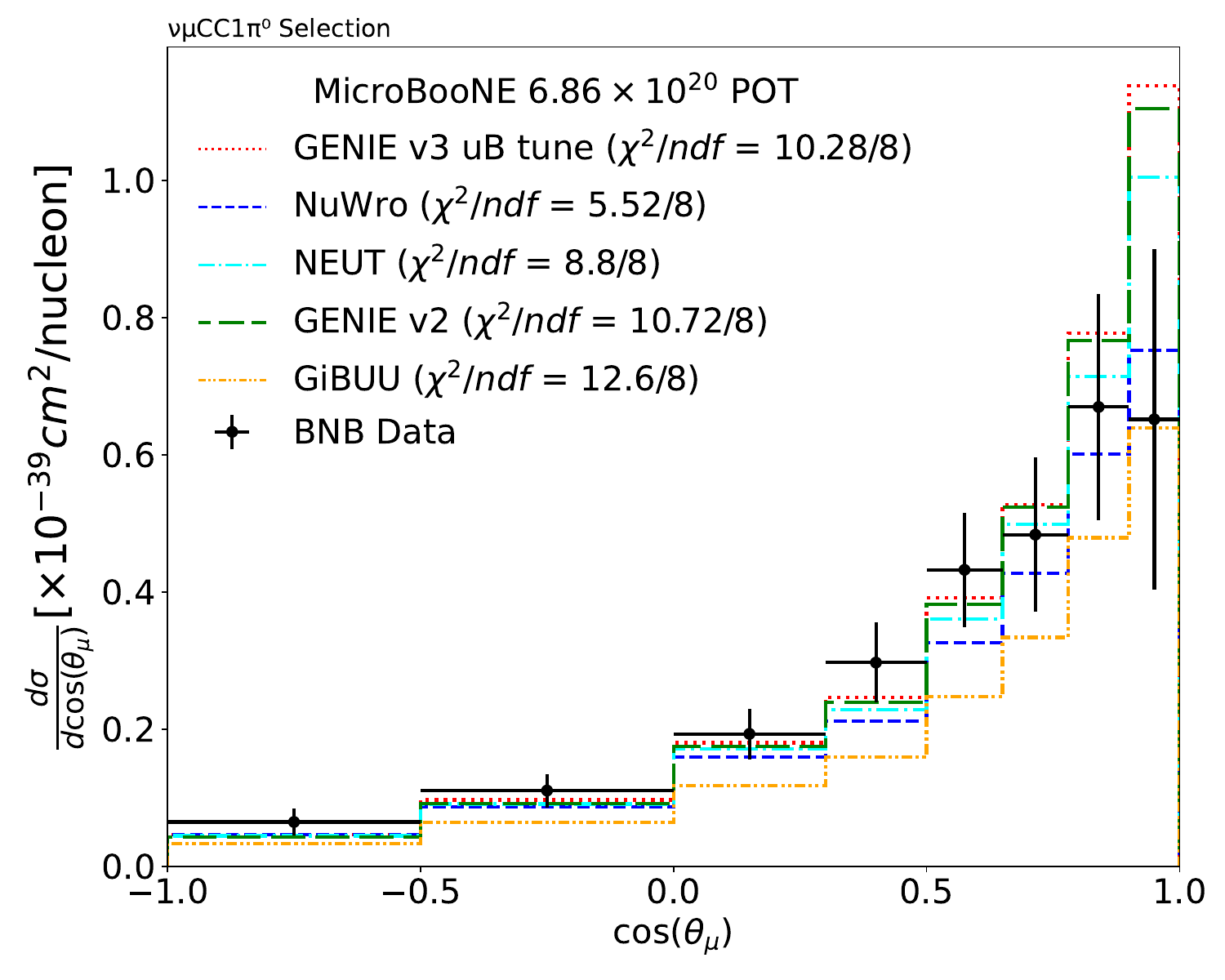}}\hfill
  \subfloat[\label{sfig:pimuanglexsec}]{
  \includegraphics[width=0.45\textwidth]{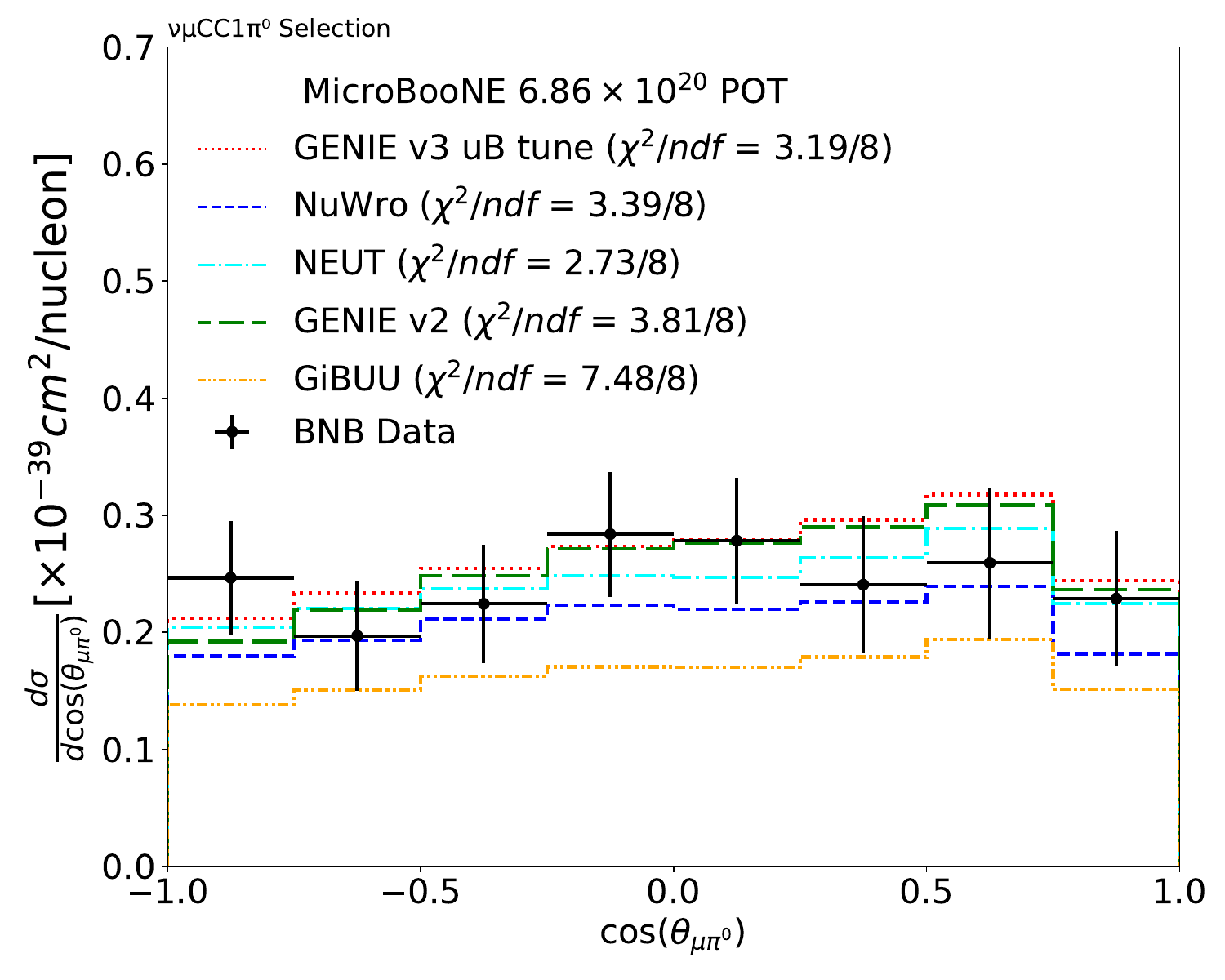}}
 \caption{\label{fig:2} Differential cross sections from several generator predictions and the unfolded data for (a) the $\pi^0$ momentum, (b) the scattering angle between the neutrino beam and outgoing $\pi^0$direction, (c) the muon momentum, (d) the scattering angle between the neutrino beam and outgoing muon direction, and (e) the opening angle between the muon and pion are shown. We quantify the agreement in terms of $\chi^2$ values, and list them in the legends. The total uncertainty on the data extracted cross section corresponding to the square root of the diagonal elements of the extracted covariance matrix is shown by the error bars. } 
\end{figure*}

\section{Cross Section Extraction}
One of the key ingredients to the cross section extraction is constructing a response matrix that accounts for the inefficiencies and limitations of the reconstruction, and maps the signal event counts in a given true bin to the observed reconstructed signal event counts after the event selection. This matrix encodes the smearing between the true and reconstructed space through the off-diagonal terms. The choice of bin widths of all five variables have been guided by the following two criteria: each bin has a minimum predicted signal event count of $\sim$50 given the limited size of the data sample after passing the $\nu_{\mu}$ CC$\pi^0$ selection, and approximately $50$\% of the selected simulated events are reconstructed in the diagonal bins in the smearing matrix. 
The differential cross-section measured in bin $i$ is given by
\begin{equation}
    \left\langle\frac{d\sigma}{dp}\right\rangle_i = \frac{\sum_j U_{ij}(N_j - B_j)}{N_{\textrm{target}} \times \phi \times (\Delta p)_i}
\end{equation}
where the unfolding matrix elements $U_{ij}$ transforms the background subtracted reconstructed events in a given bin $j$ to true bin $i$, and $(\Delta p)_i$ is the width of bin $i$. For a variable $p$, $N_j$ and $B_j$ are the numbers of selected data and background events in bin $j$ respectively. The variables $\phi$ and $N_{\textrm{target}}$ correspond to the flux and the number of argon targets in the fiducial volume. The unfolding procedure involves inversion of the response matrix, which can lead to an unfolded distribution with large variance in the true variable space. This is taken into account by introducing regularization conditions. We extract the cross sections using the Wiener-SVD unfolding method~\cite{Wiener_SVD} with a regularization approach corresponding to a first-order derivative. The effect of regularization is quantified by an additional smearing matrix and is applied to the generator predictions before comparing with the unfolded spectrum. An unfolded result can be directly compared to various theoretical predictions as it corrects for detector efficiency and smearing. The additional smearing matrices used for the extraction of the results presented in Fig.~\ref{fig:2} are provided in the supplemental material~\cite{supp_mat}. %is applied to the unfolded spectrum before comparing with various generator predictions.  
We validate the Wiener-SVD unfolding technique by performing fake data studies before unfolding the selected BNB data events. 
The BNB data extracted cross sections are shown in Fig. \ref{fig:2} along with a series of generator predictions; the MicroBooNE tuned version of GENIE $\text{v3.0.6}$ (GENIE v3 uB tune)~\cite{PhysRevD.105.072001}, GENIE $\text{v2.12.2}$ (GENIE v2)~\cite{ANDREOPOULOS201087, andreopoulos2015genie}, NuWro $\text{19.02.1}$ (NuWro)~\cite{GOLAN2012499, PhysRevC.86.015505}, Neut $\text{v5.4.0}$ (NEUT)~\cite{Hayato:2009zz, NEUT} and GiBUU 2023 (GiBUU)~\cite{BUSS20121}. The difference in prediction from these generators comes from the different underlying models. Several MicroBooNE publications describe these models in detail~\cite{PhysRevLett.125.201803, PhysRevD.105.L051102, PhysRevD.102.112013}. The generators have different initial state nuclear models (GENIE v2 uses a relativistic Fermi gas, while the others use a local Fermi gas) and resonant pion production models. The NuWro generator implements the Adler-Rarita-Schwinger formalism~\cite{PhysRevD.80.093001} to explicitly calculate the $\Delta (1232)$ resonance, and the non-resonant background is estimated using a quark-parton model. GENIE v2 uses the Rein and Sehgal model~\cite{REIN198179}, while NEUT and the tuned version of GENIE v3 follow the Berger and Sehgal approach~\cite{10.1063/1.3274164,PhysRevD.76.113004,PhysRevD.79.079903}. GiBUU employs the MAID analysis~\cite{Mosel_2019} for modeling the RES interactions. The different generators also differ in their treatmeant of final state interactions (FSI). NuWro employs intranuclear cascade models~\cite{PhysRevC.86.015505}, NEUT uses the FSI cascade approach with nuclear medium corrections for pions~\cite{SALCEDO1988557}. GiBUU uses numerical solution from the Boltzmann-Uehling-Uhlenbeck equation to model the intranuclear hadron transport. As a result, the initial state and the FSI effects are described in a consistent nuclear potential. The GENIE generator versions used here applies the hA FSI model~\cite{10.1063/1.3661588,10.1063/1.4931864}. The differences among various generators are presented in a tabular format in~\cite{acero2021measurement}.  
\begin{table}[h]
    \begin{ruledtabular}
    \begin{tabular}{cccccc}
    Generator & $p_{\pi^0}$ & $\cos(\theta_{\pi^0})$ & $p_{\mu}$ & $\cos(\theta_{\mu})$ & $\cos(\theta_{\mu\pi^0})$\\
      \hline
         GENIE v3(tuned) & 0.054 & 0.859 & 0.0063 & 0.246 & 0.922 \\
         NuWro & 0.0079 & 0.748 & 0.0007 & 0.701 & 0.908 \\
         NEUT & 0.059 & 0.788 & 0.0012 & 0.360 & 0.950 \\
         GENIE v2 & 0.0002 & 0.964 & 0.0046 & 0.218 & 0.874 \\
         GiBUU & 0.008 & 0.465 & 0.0003 & 0.126 & 0.486 \\
    \end{tabular}
    \end{ruledtabular}
    \caption{ P-values comparing unfolded data and generator predictions given in Fig.~\ref{fig:2}.}
    \label{tab:my_label}
\end{table}
The comparison between generator predictions and the data extracted cross sections is quantified in terms of $\chi^2$ over number of degrees of freedom, included in the legends in Fig.~\ref{fig:2} and p-values, presented in Table \ref{tab:my_label}. The results are discussed in detail in the following section. 

\section{Results}

Figure~\ref{sfig:pi0momxsec} shows the flux averaged differential cross section as a function of $\pi^0$ momentum. Pions interacting with the nuclear medium can shift the momentum to lower values, therefore contributing to the buildup of the peak. The generator predictions underestimate the data extracted cross section in the peak (around 100-200 MeV) and tend to have a better agreement at higher momentum. At low momentum, NuWro predictions are much lower than the data, but are comparable with other generator predictions as the momentum increases. Predictions from NEUT and NuWro become very similar after the peak. As the FSI effects play a role in shifting the distribution towards lower momenta, the comparison indicates an underestimation of pion FSI considered in NuWro among the generators that use the same initial nuclear model. GiBUU underestimates the cross section in the FSI dominated range as well (comparisons of various GiBUU generator model choices can be found in~\ref{GiBUU}). The poorest agreement between prediction and data come from GENIE v2 with a $\chi^2/ndf = 30.044/8$ (p-value = 0.0002). In the lower momentum region, it underestimates the data and in the higher momentum bins this prediction is enhanced. 

Figure~\ref{sfig:pi0anglexsec} presents the flux averaged differential cross section in $\pi^0$ scattering angle. The data and the generator predictions agree within the uncertainties, except GiBUU predictions underestimating the cross section in the forward angles. The $\chi^2/ndf$ values indicate that a similar level of agreement is achieved by all other generators for the $\pi^0$ production angle. 

Figure~\ref{sfig:mumomxsec} shows the flux averaged differential cross section in muon momentum. The data and generator comparison follows a similar trend as the $\pi^0$ momentum with suppressed generator predictions in the lower momenta (around 200 - 400 MeV) whereas in the higher momenta the scenario is inverted and the generator predictions are higher than the data. GiBUU has the poorest agreement with data with a $\chi^2/ndf = 28.92/8$ (p-value = 0.0003). The range of $\chi^2/ndf$ values is broader for the $\pi^0$ momentum in Fig.~\ref{sfig:pi0momxsec} than for the muon momentum. 

Figure~\ref{sfig:muanglexsec} presents the flux averaged differential cross section in muon scattering angle. The generator predictions are enhanced compared to cross section extracted from data at forward angles $\cos(\theta_{\mu}) > 0.9 $, while there is a similar level of agreement for $\cos(\theta_{\mu}) < 0.9 $. The discrepancy in the forward angle cannot be explained by the systematic uncertainties and indicates shortcomings of the generator models. The forward angle corresponds to low momentum transfer events which were previously observed to not be well reproduced by models in MINERvA~\cite{PhysRevD.96.072003}, and MiniBooNE measurements~\cite{PhysRevD.83.052009}. Previous CCQE-enhanced MicroBooNE cross section measurements have demonstrated good agreement with models with low momentum transfer suppression implemented~\cite{PhysRevD.108.053002}. Figure~\ref{sfig:muanglexsec} indicates that a similar treatment in the RES interactions can improve the agreement. An improvement in $\chi^2$ values for $\pi^0$ production has been reported by the MINERvA collaboration~\cite{PhysRevD.100.072005} after tuning the GENIE pion production model with MINERvA data. At the beam energies of this measurement, muons and pions are dominantly produced at forward angles in the laboratory frame due to the Lorentz boost. The dominant interaction in the forward angle for both muons and pions arise from RES interactions. The measured differential cross section as a function of the muon scattering angle $\cos\theta_{\mu}$ is in reasonable agreement with all models. NuWro gives the best prediction due to its lower normalization. Figure~\ref{sfig:pimuanglexsec} shows the extracted cross section in the muon pion opening angle, where all generators demonstrate a good data-MC agreement.

\section{Conclusions}
In this paper, we discuss the event selection strategies, sources and estimation of systematic uncertainties, and Wiener-SVD unfolding technique for a $\nu_{\mu}$ CC1$\pi^0$ single-differential cross section measurement in outgoing muon and pion kinematic variables. The dominant sources of systematic uncertainties come from detector response systematics and neutrino flux. We compare the measured cross sections with predictions from several neutrino generators and report good agreement for muon and pion scattering angles, except for an overprediction by generators at muon forward angles. This suggests that there is scope for improvement in the generator models for the resonant interactions. An underprediction by generators in the medium momentum ranges, $200 - 400$ MeV for muons and $100 - 200$ MeV for pions show a similar  trend as previously reported by the MINERvA~\cite{PhysRevD.96.072003} and MiniBooNE collaborations~\cite{PhysRevD.83.052009} with similar signal definitions. 
 
While this measurement is systematics limited, future larger statistics measurements would allow for a double differential cross section extraction which may help further improve modeling of resonance processes.

\section{Acknowledgements}

This document was prepared by the MicroBooNE collaboration using the resources of the Fermi National Accelerator Laboratory (Fermilab), a U.S. Department of Energy, Office of Science, HEP User Facility. Fermilab is managed by Fermi Research Alliance, LLC (FRA), acting under Contract No. DE-AC02-07CH11359. MicroBooNE is supported by the following: the U.S. Department of Energy, Office of Science, Offices of High Energy Physics and Nuclear Physics; the U.S. National Science Foundation; the Swiss National Science Foundation; the Science and Technology Facilities Council(STFC), part of the United Kingdom Research and Innovation; the Royal Society (United Kingdom); and the UK Research and Innovation (UKRI) Future Leaders Fellowship. Additional support for the laser calibration system and cosmic ray tagger was provided by the Albert Einstein Center for Fundamental Physics, Bern, Switzerland. We also acknowledge the contributions of technical and scientific staff to the design, construction, and operation of the MicroBooNE detector as well as the contributions of past collaborators to the development of MicroBooNE analyses, without whom this work would not have been possible. For the purpose of open access, the authors have applied a Creative Commons Attribution (CC BY) public copyright license to any Author Accepted Manuscript version arising from this submission.
\appendix
\section{GiBUU model comparisons \label{GiBUU}}
\begin{figure}
    \subfloat[\label{sfig:pi0momxsec}]{
  \includegraphics[width=0.45\textwidth]{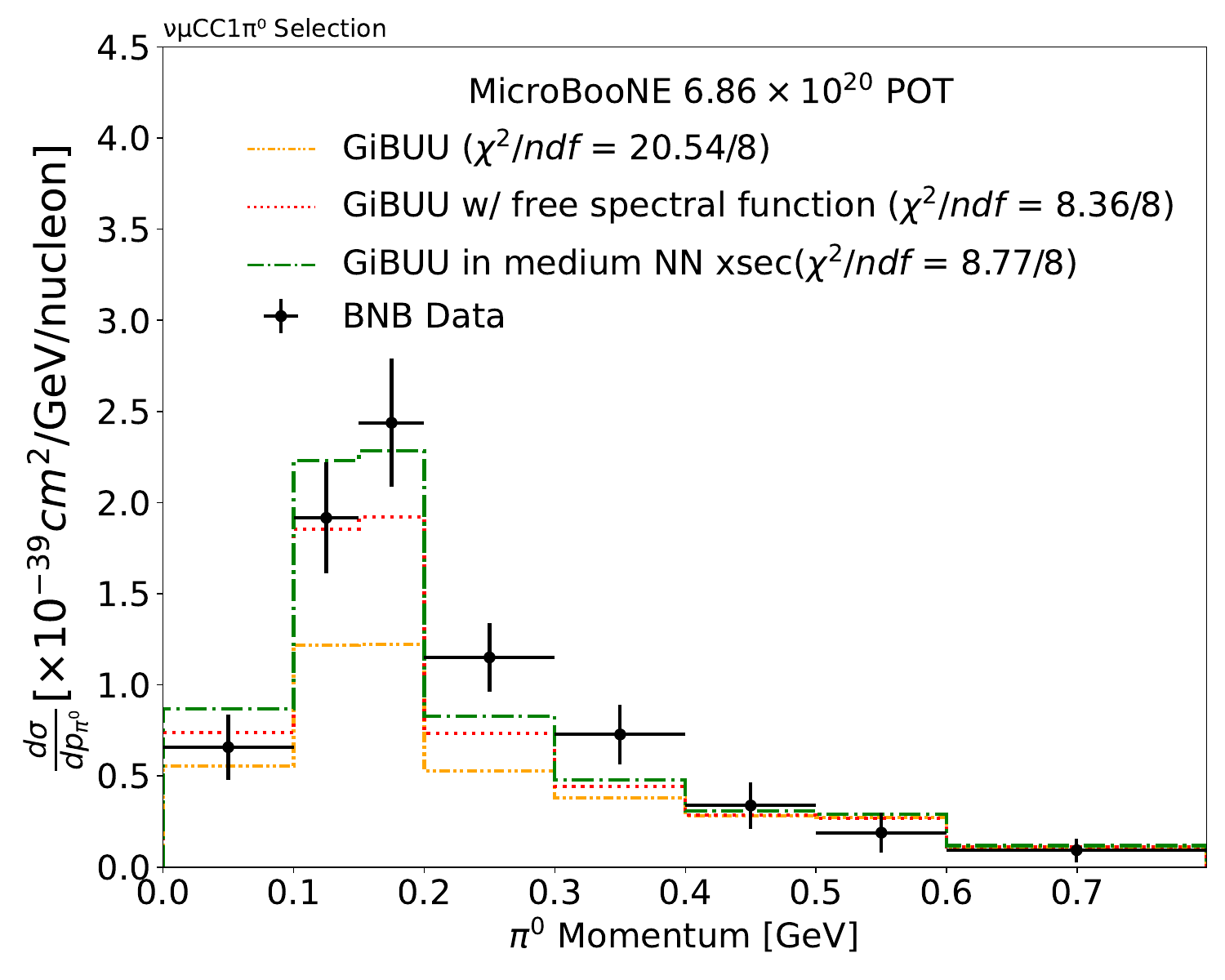}} \vfill
  \subfloat[\label{sfig:pimuanglexsec}]{
  \includegraphics[width=0.45\textwidth]{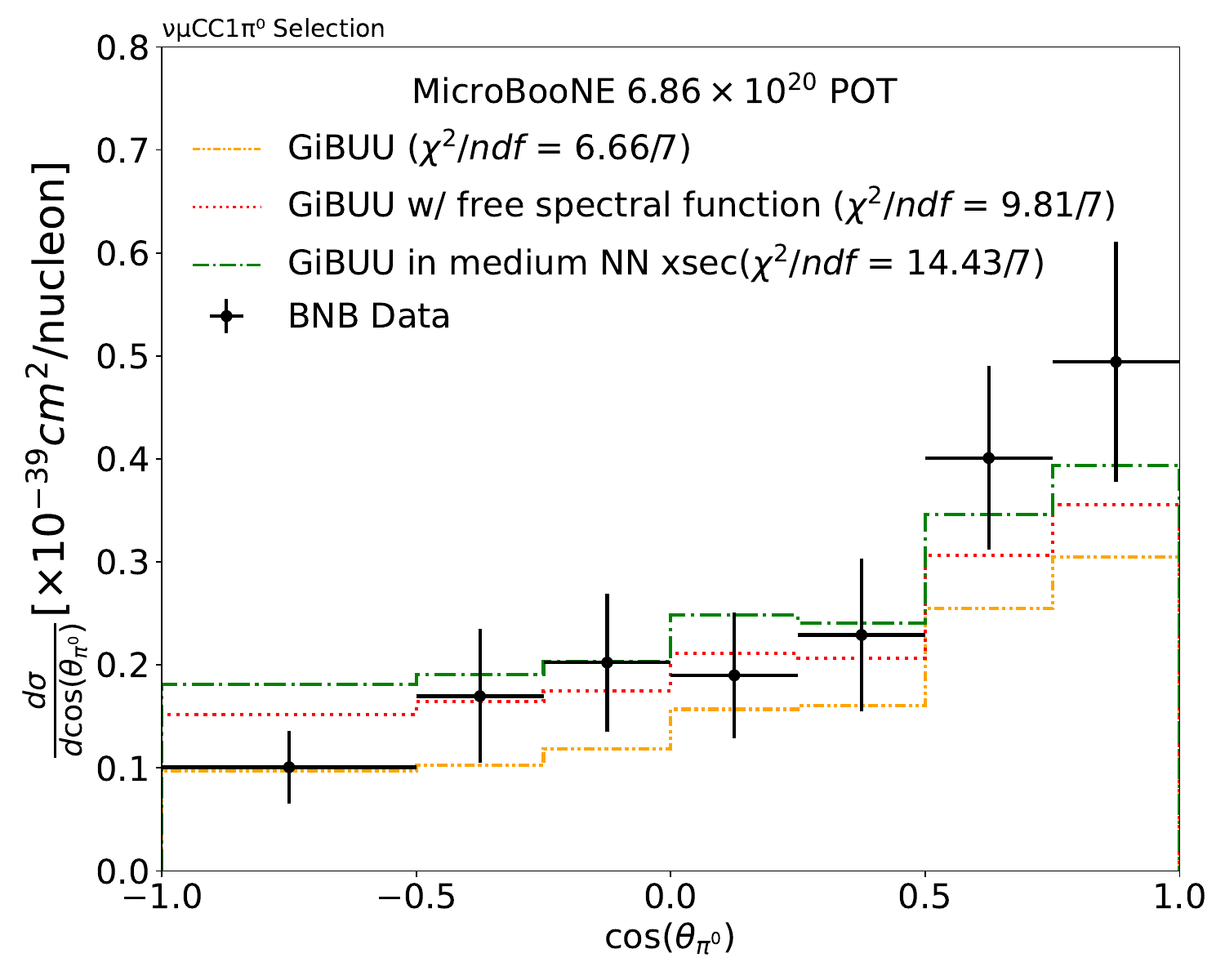}}\vfill
  \caption{Comparing differential cross sections from various GiBUU model choices and the unfolded data for (a) the $\pi^0$ momentum, (b) the scattering angle between the neutrino beam and outgoing $\pi^0$direction. We quantify the agreement in terms of $\chi^2$ values, and list them in the legends.}\label{fig:gibuu}
\end{figure}

\begin{figure*}
  \subfloat[\label{sfig:pimuanglexsec}]{
  \includegraphics[width=0.45\textwidth]{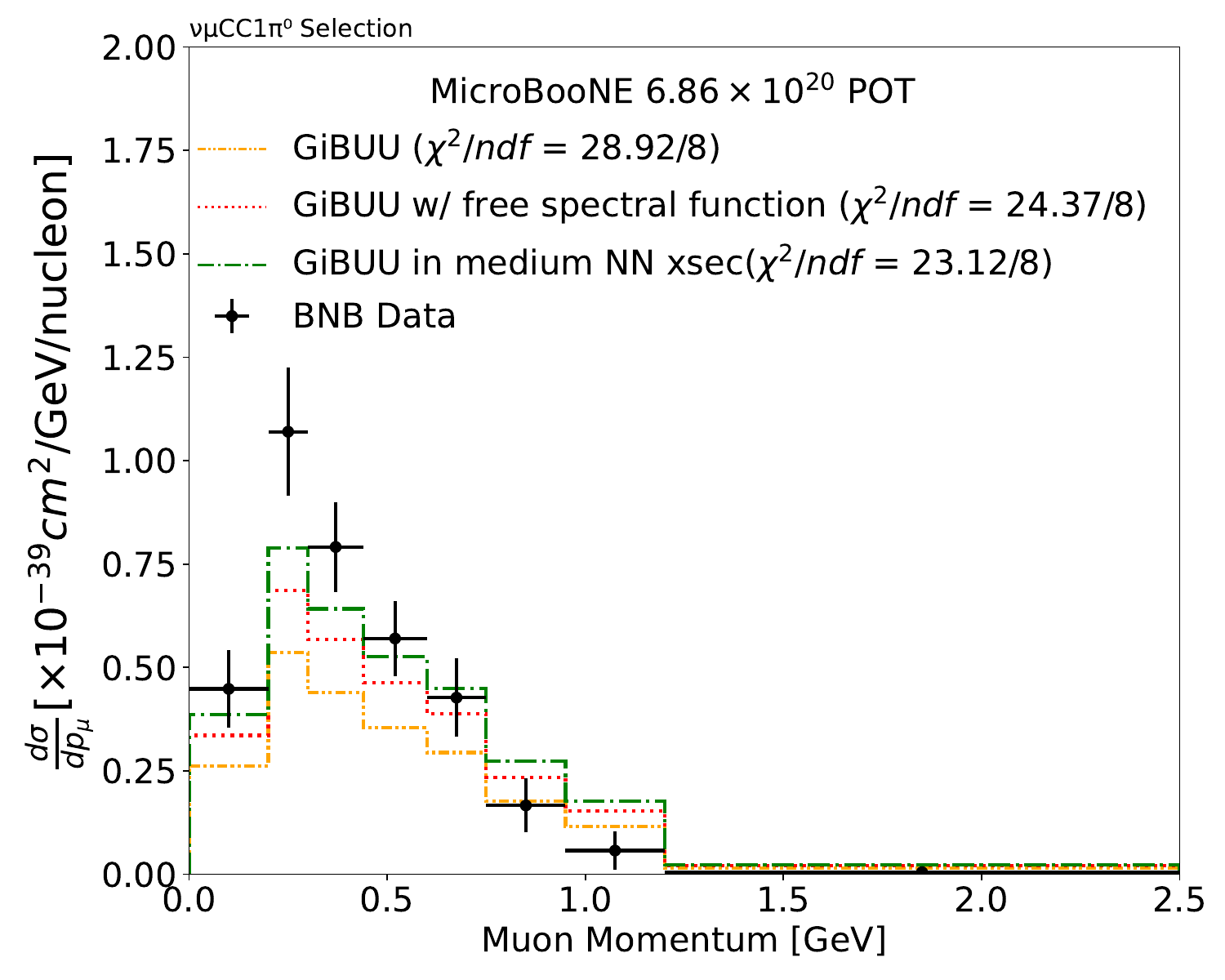}}\hfill
  \subfloat[\label{sfig:pimuanglexsec}]{
  \includegraphics[width=0.45\textwidth]{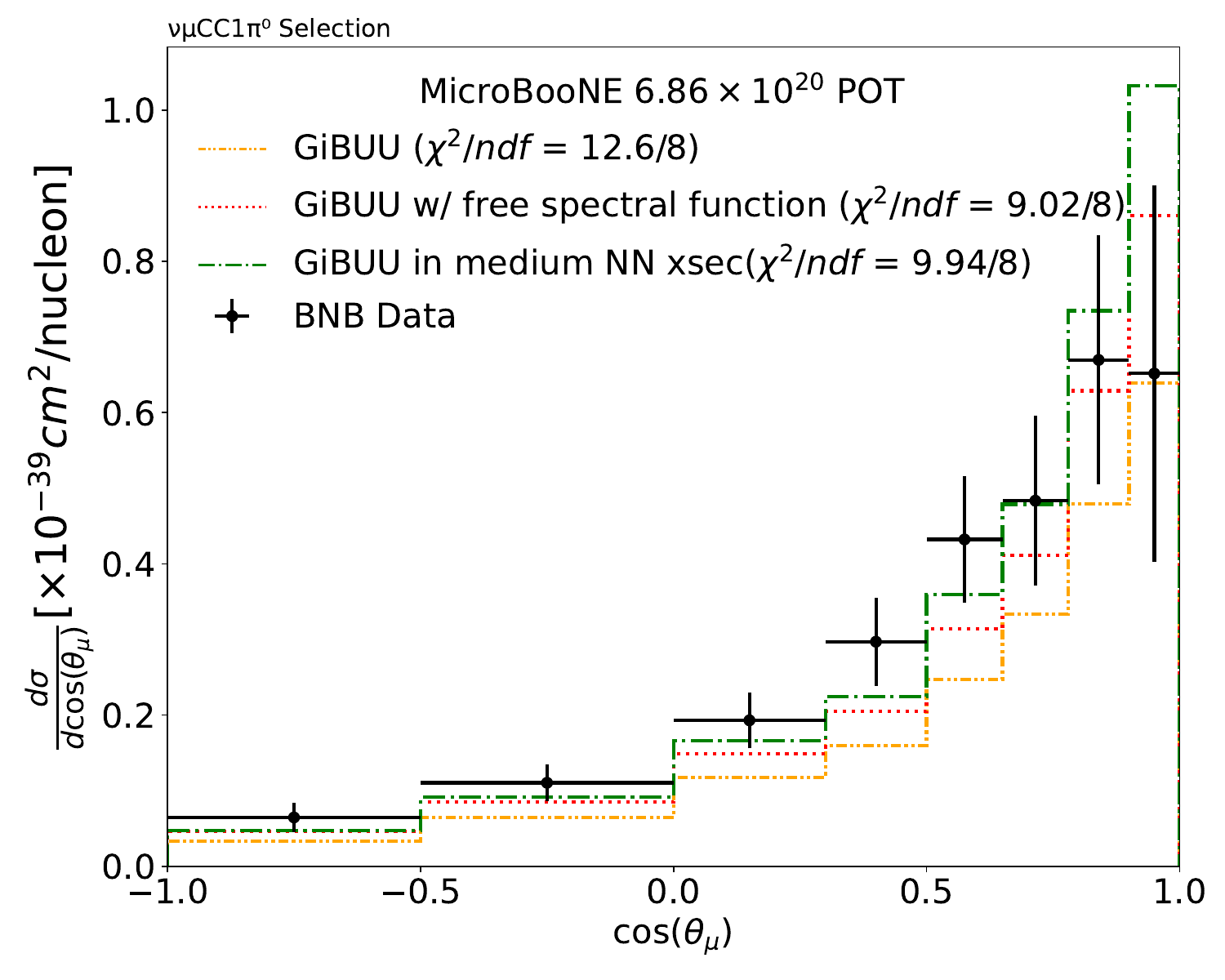}}\hfill
  \subfloat[\label{sfig:pimuanglexsec}]{
  \includegraphics[width=0.45\textwidth]{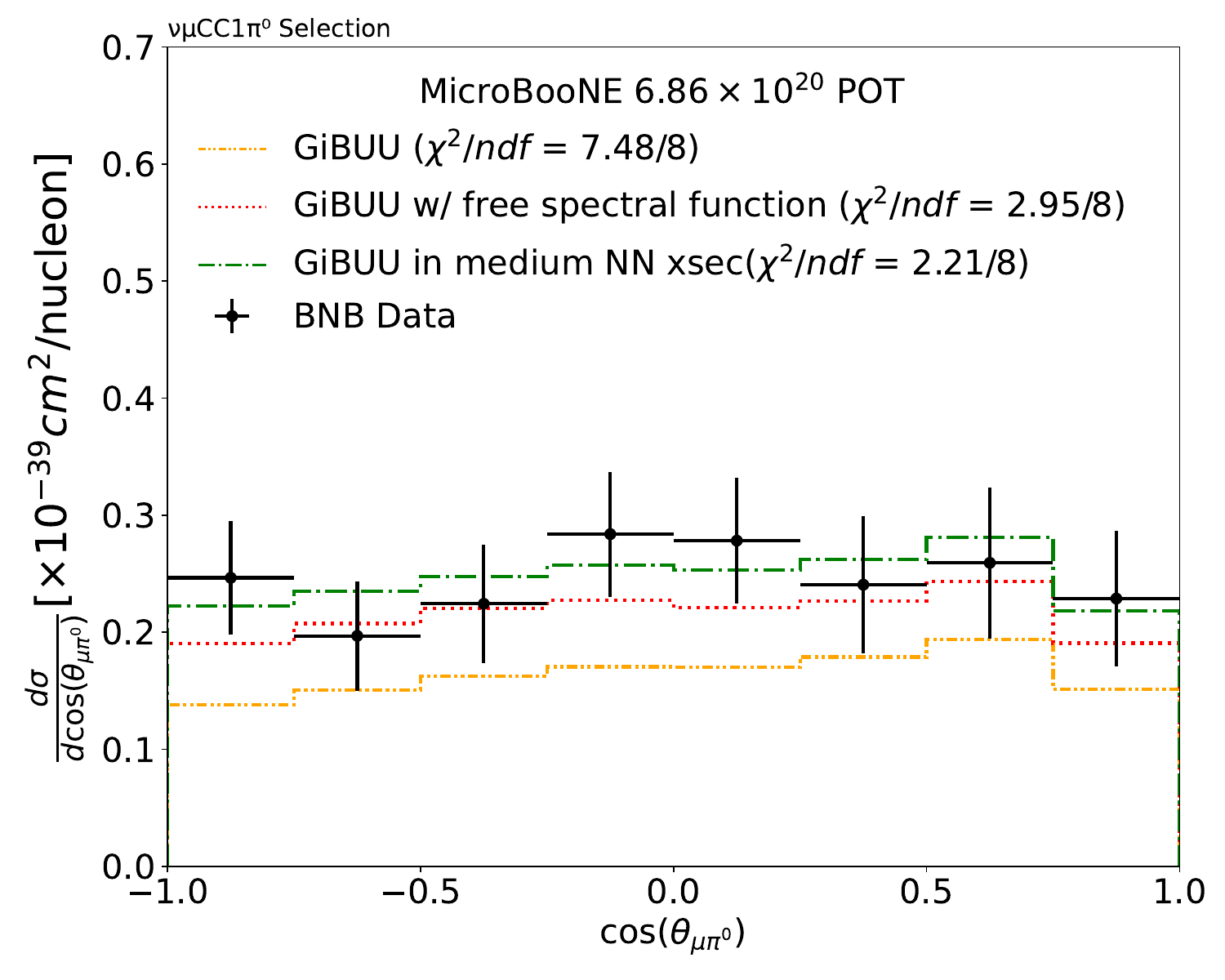}}\hfill
 \caption{Comparing differential cross sections from various GiBUU model choices and the unfolded data for (a) the muon momentum (b) the scattering angle between the neutrino beam and outgoing muon direction and (c) the muon pion opening angle. We quantify the agreement in terms of $\chi^2$ values, and list them in the legends.}\label{fig:gibuu1}
\end{figure*}

We compare three model choices for the GiBUU generator prediction in Figs.~\ref{fig:gibuu} and~\ref{fig:gibuu1}. The default version of GiBUU 2023 generator uses the Valencia-Oset collisional broadening for the $\Delta$ resonance~\cite{BUSS20121, Mosel_2019}. We compare the performance of GiBUU generator predictions by using the free spectral function (which can be activated by setting mediumSwitch = false in the relevant GiBUU jobcard). Apart from the free spectral function prediction, we also report the predictions with nucleon-nucleon (NN) cross sections calculated in medium (flagInMedium = true and mediumSwitch = false)~\cite{PhysRevC.49.566, PhysRevC.48.1702, PhysRevC.91.014901}. The default version of GiBUU 2023 uses the NN cross sections in vacuum while evaluating the FSI. Figures~\ref{fig:gibuu} and~\ref{fig:gibuu1} show the three GiBUU model comparisons in the five kinematic variables of interest in this analysis.

\clearpage
\newpage
\bibliography{bibliography}% Produces the bibliography via BibTeX.

\end{document}